\documentclass[
    prd,
    superscriptaddress,
    reprint,
    preprintnumbers,
    amsmath,
    amssymb,
    aps,
    nofootinbib,
]{revtex4-2}
\usepackage[T1]{fontenc}

\usepackage{hyperref}

\usepackage{amsmath}
\usepackage{amsfonts,amssymb}
\usepackage{dcolumn}

\newcommand*{\smallrel}[2][.8]{%
  \mathrel{\mathpalette{\smallrel@{#1}}{#2}}%
}
\makeatother

\def\({\left(}
\def\){\right)}

\newcommand\rmd { {\rm d} }

\newcommand{\dd}{\ensuremath \mathrm{d}}

\usepackage{tikz}
\usepackage{slashed}
\usepackage{pgfplots}
\usepackage{mathtools}
\usepackage{siunitx}
\usepackage{physics}
\usepackage[compat=1.1.0]{tikz-feynman}
\tikzset{graviton/.style={decorate, decoration={snake, amplitude=.6mm, segment length=1.5mm, pre length=.3mm, post length=.3mm}, double}}

\usepackage{xcolor}
\xdefinecolor{redviolet}{RGB}{228,38,207}
\definecolor{pastelgreen}{RGB}{210, 245, 210}
\definecolor{2gFSR}{RGB}{117,20,124}
\definecolor{2gFSRlight}{RGB}{210,179,213}
\definecolor{2gISR}{RGB}{55,126,33}
\definecolor{2gISRlight}{RGB}{185,215,181}
\definecolor{1gISR}{RGB}{202,102,39}
\definecolor{1gISRlight}{RGB}{237,208,182}
\definecolor{blueblob}{HTML}{4C72B0}
\xdefinecolor{brickred}{RGB}{220,60,66}
\xdefinecolor{bluscuro}{RGB}{51,51,179}

\usepackage{fontawesome5}
\usepackage{amssymb}

\begin{document}
\preprint{RBI-ThPhys-2026-20}
\title{Next-to-leading order corrections to radiative pion pair production via dispersion relations}

\author{Carlo M. Carloni Calame}
 \email{carlo.carloni.calame@pv.infn.it}
\affiliation{INFN, Sezione di Pavia, Via A. Bassi 6, 
27100, Pavia, Italy}

\author{Marco Ghilardi}
 \email{marco.ghilardi01@universitadipavia.it}
\affiliation{Dipartimento di Fisica ``Alessandro Volta'', Universit\`a di Pavia, Via A. Bassi 6, 27100, Pavia, Italy}
\affiliation{INFN, Sezione di Pavia, Via A. Bassi 6, 
27100, Pavia, Italy}

\author{Andrea Gurgone}
 \email{andrea.gurgone@irb.hr}
\affiliation{Division of Theoretical Physics, Ru{\dj}er Bo\v{s}kovi\'{c} Institute, Bijeni\v{c}ka Cesta 54, 10000, Zagreb, Croatia}

\author{Guido Montagna}
 \email{guido.montagna@unipv.it}
\affiliation{Dipartimento di Fisica ``Alessandro Volta'', Universit\`a di Pavia, Via A. Bassi 6, 27100, Pavia, Italy}
\affiliation{INFN, Sezione di Pavia, Via A. Bassi 6, 
27100, Pavia, Italy}

\author{Mauro Moretti}
 \email{mauro.moretti@fe.infn.it}
\affiliation{Dipartimento di Fisica e Scienze della Terra, Universit\`a di Ferrara, Via Saragat 1, 44122, Ferrara, Italy}
\affiliation{INFN, Sezione di Ferrara, Via Saragat 1, 44122, Ferrara, Italy}

\author{Oreste Nicrosini}
 \email{oreste.nicrosini@pv.infn.it}
\affiliation{INFN, Sezione di Pavia, Via A. Bassi 6, 
27100, Pavia, Italy}

\author{Fulvio Piccinini}
 \email{fulvio.piccinini@pv.infn.it}
\affiliation{INFN, Sezione di Pavia, Via A. Bassi 6, 
27100, Pavia, Italy}
\affiliation{INFN, Galileo Galilei Institute for 
Theoretical Physics, Largo E. Fermi 2, 50125, Firenze, Italy}
\author{Francesco P. Ucci}
 \email{francesco.ucci@pv.infn.it}
\affiliation{Dipartimento di Fisica ``Alessandro Volta'', Universit\`a di Pavia, Via A. Bassi 6, 27100, Pavia, Italy}
\affiliation{INFN, Sezione di Pavia, Via A. Bassi 6, 
27100, Pavia, Italy}

\date{\today}

\begin{abstract}
We compute the next-to-leading order corrections to the radiative return process $e^+ e^- \to \pi^+ \pi^- \gamma$ using a dispersive representation of the pion form factor for the calculation of loop integrals. We compare our results with those of the factorised scalar QED approach, as well as with previous predictions obtained by us in the generalised vector meson dominance model.
We show the numerical impact of the structure-dependent corrections on various observables of interest for radiative return experiments at flavour factories. Following recent input from the literature, we include in our calculation also corrections due to neutral pseudoscalar poles and we provide a first estimate of such contributions.
We also investigate additional mechanisms contributing to final-state radiation at center-of-mass energies around the $\phi$-meson resonance, such as radiative $\phi$ decays, presenting numerical results that are relevant for the KLOE experiment.
These novel features are implemented in the Monte Carlo event generator \textsc{BabaYaga@NLO}, which can now be used to evaluate the impact of the modelling of pion-photon interaction in radiative return measurements. 

\end{abstract}

\maketitle

\section{Introduction}
\label{sec:introduction}

The hadronic cross section in $e^+ e^-$ annihilation at low energies is a key quantity for the data-driven computation of the hadronic 
vacuum polarisation (HVP) correction to the anomalous 
magnetic moment of the muon, ${a_\mu = (g - 2)_\mu / 2}$, as well as of the running of the 
electromagnetic coupling constant~\cite{Jegerlehner:2017gek,Colangelo:2018mtw,Keshavarzi:2019abf,Benayoun:2019zwh,Davier:2019can,Aoyama:2020ynm,Keshavarzi:2024wow,Aliberti:2025beg,Keshavarzi:2026xwi}.

For the calculation of the HVP contribution to $a_{\mu}$, the dominant channel is the 
process of two pion production $e^+ e^- \to \pi^+ \pi^-$, that represents the main source of theoretical uncertainty. This process is measured at $e^+ e^-$ colliders at the GeV scale (flavour 
factories) in energy scan~\cite{CMD-3:2023alj,CMD-3:2023rfe,Achasov:2006vp,SND:2020nwa,CMD-2:2001ski,CMD-2:2005mvb,CMD-2:2006gxt,1989321}
and radiative return~\cite{BaBar:2012bdw,BESIII:2015equ,KLOE:2004lnj,KLOE:2008fmq,KLOE:2010qei,KLOE:2012anl,KLOE-2:2016mgi,KLOE-2:2017fda} experiments. 
From these measurements, one can extract the time-like electromagnetic pion form factor $F_\pi(q^2)$, that provides information about the non-perturbative structure of the pion.

The available measurements of $F_\pi (q^2)$ show some tensions that affect the data-driven evaluations of
the HVP contribution to $a_{\mu}$. Actually, the predictions of $a_\mu$ based on the data-driven method are not compatible with the muon $g-2$ experimental world average~\cite{Muong-2:2025xyk}, except for the CMD-3 determination~\cite{Aliberti:2025beg}. New measurements of $F_\pi (q^2)$ with sub-percent precision are ongoing or planned in the near future, by using 
both the energy scan and the radiative return method~\cite{SND_Phipsi2026,BELLE_orsay2025,KLOE_orsay2025,BES3_orsay2025,Zhang:2026jvb,Polat:2026ysh}. This experimental effort requires parallel theoretical progress to keep all the relevant sources of systematic uncertainty under control. Actually, the forthcoming measurements stimulated updated comparisons between the predictions of the existing Monte Carlo (MC) codes for experiments at flavor factories~\cite{Aliberti:2024fpq}, as well as the construction of new precise tools~\cite{Banerjee:2020rww} and the development of new versions of existing generators with improved theoretical accuracy~\cite{Budassi:2024whw,Budassi:2026lmr,Sherpa:2024mfk,Price:2025fiu}. 

In this general context, a topic of high current interest is the calculation of radiative corrections to processes with pions in the final state. Indeed, in a series of papers~\cite{Ignatov:2022iou,Colangelo:2022lzg,Budassi:2024whw,Gurgone:2025cci,Fang:2025mhn,Colangelo:2025ivq,Monnard:2021pvm,Flores-Baez:2025nra}, it was recently shown that the insertion of the composite 
structure of the pion in one-loop computations is essential to obtain sensible predictions for various observables 
of experimental interest. In particular, for the
energy scan process $e^+ e^- \to \pi^+ \pi^-$, the structure-dependent corrections are crucial to obtain predictions 
for the forward-backward (or charge) asymmetry in agreement with the data of the CMD-3 collaboration~\cite{Ignatov:2022iou,Colangelo:2022lzg,Budassi:2024whw,Fang:2025mhn}.

To take into account the internal structure of the pion in next-to-leading order (NLO) calculations, it is necessary to go beyond the naive factorised scalar QED (F$\times$sQED) approach~\cite{Rodrigo:2001jr,Rodrigo:2001kf,Kuhn:2002xg,Czyz:2002np,Czyz:2003ue,Campanario:2019mjh,Budassi:2024whw,Budassi:2026lmr}, where 
the point-like scalar QED (sQED) amplitudes are multiplied
by an overall pion form factor, which is evaluated
at an appropriate virtuality to ensure the cancellation of infrared (IR) divergences~\cite{Tracz:2018nkj}. This amounts to calculate the so-called structure-dependent corrections, as it requires to introduce analytically the pion form factor as a function of the loop momentum in virtual amplitudes entering the process under consideration. Such challenging task can be performed according to two independent methods: the generalised
vector meson dominance (GVMD) model~\cite{Ignatov:2022iou,Budassi:2024whw,PetitRosas:2026iuq,CarloniCalame:2026vfc}
and the so-called FsQED approach~\cite{Colangelo:2014dfa,Colangelo:2015ama,Colangelo:2022lzg,Budassi:2024whw}. In~\cite{Colangelo:2014dfa,Colangelo:2015ama}, FsQED is 
defined as the multiplication of the sQED amplitude by the pion form factors evaluated at the photon momentum transfers in loop integrals, where $F_\pi(q^2)$ is written by means of its dispersive
representation~\footnote{For the subprocess $\gamma^*\gamma^*\to\pi^+\pi^-$ entering the calculation of $e^+e^-\to\pi^+\pi^-$ and of the Hadronic Light by Light (HLbL) contribution to $a_\mu$, FsQED is equivalent to isolate the charged-pion pole~\cite{Colangelo:2015ama,Colangelo:2022lzg}}. In the present work, we apply this procedure to all the loop topologies of $e^+e^-\to\pi^+\pi^-\gamma$, thus including the corrections beyond 
F$\times$sQED in all diagrams, irrespectively of which charged leg emitted the real photon.

The GVMD model was adopted in~\cite{Ignatov:2022iou,Budassi:2024whw} for the computation of the NLO corrections to the process of two pion production in $e^+ e^-$ annihilation. These calculations were implemented in the MC programs \textsc{MCGPJ}~\cite{Ignatov:2022iou,Arbuzov:1997je} and \textsc{BabaYaga@NLO}~\cite{Budassi:2024whw, Balossini:2006wc,Balossini:2008xr}. The application of the FsQED approach to $e^+ e^- \to \pi^+ \pi^-$ was
presented in~\cite{Colangelo:2022lzg}. The NLO structure-dependent corrections 
to the energy-scan process, as computed
according to dispersive representations of the pion form factor, are available in the codes \textsc{BabaYaga@NLO}~\cite{Budassi:2024whw} and 
\textsc{McMULE}~\cite{Fang:2025mhn}.

Recently, the GVMD model was also applied to the calculation of one-loop virtual corrections to the radiative return process $e^+ e^- \to \pi^+ \pi^- \gamma$ in~\cite{CarloniCalame:2026vfc,PetitRosas:2026iuq}. The computation of~\cite{PetitRosas:2026iuq} was interfaced to the event generator \textsc{Phokhara}, while the calculation of~\cite{CarloniCalame:2026vfc} was implemented in \textsc{BabaYaga@NLO}. The application of the FsQED approach to radiative return is still missing~\footnote{After the submission of this work, an independent calculation of structure-dependent corrections to radiative pion-pair production using a dispersive representation of the pion form factor appeared in~\cite{Fang:2026naz}.}.

The goal of this work is to fill this gap in the literature, by using the FsQED approach to take into account the non-perturbative nature of the pion in the computation of the NLO corrections to the radiative process $e^+ e^- \to \pi^+ \pi^- \gamma$. We aim to study how the FsQED predictions differ from the results of the F$\times$sQED approximation, in order to quantify the uncertainty associated with the modelling of the pion-photon interaction in the computation of the 
NLO corrections to radiative return. Moreover, a further goal is to compare the FsQED results with those obtained 
by us in~\cite{CarloniCalame:2026vfc} according to the GVMD model, to assess to what extent the predictions for the structure-dependent correction in the two model 
are in agreement. In this respect, our work extends the results obtained in~\cite{Budassi:2024whw} from energy scan to radiative return and completes the treatment of the structure-dependent corrections 
in \textsc{BabaYaga@NLO}. Following~\cite{Aliberti:2024fpq,Hof_turin}, we also investigate corrections beyond FsQED that could be enhanced by $P$-wave rescattering and we provide a first estimate of the numerical impact of such corrections by calculating the neutral pseudoscalar-pole contributions, as suggested in~\cite{Hoferichter_private}. For completeness, we also study the contribution due to different mechanisms that are necessary for an appropriate description of the $\pi^+ \pi^-\gamma$ final state for center-of-mass (c.m.) energies around the $\phi$-meson resonance and we show predictions of interest for comparison with KLOE data.

The paper is organised as follows. In Sec.~\ref{sec:calculation},
we describe the details of the calculation of the one-loop structure-dependent
corrections. 
The modelling of final-state radiation (FSR) beyond 
F$\times$sQED around the $\phi$ resonance is addressed in Sec.~\ref{sec:radphi}. In Sec.~\ref{sec:numres},
we show numerical results for experimentally relevant differential cross sections 
by using realistic
event selection criteria, inspired by radiative return experiments at flavour factories.
In Sec.~\ref{sec:conclusion} we summarise the main conclusions of 
our work.

\begin{figure}
\begin{minipage}{0.2\textwidth}
\begin{tikzpicture}[scale=0.5, transform shape]
\begin{feynman}[baseline=(a.b),small]
    \vertex (a) at (-2, -2) ;
    \vertex (b) at (-1, -1) ;
    \vertex (u) at (-1, 0) ;
    \vertex (p) at (-2, 0) ;
    \vertex (c) at (-1, 1) ;
    \vertex (d) at (-2, 2) ;
    \vertex [dot] (e) at (2, -2) ;
    \vertex (f) at (1, -1) ;
    \vertex (g) at (1, 0) ;
    \vertex [dot] (h) at (2, 2) ;
    \vertex (y) at (0, 2) ;
    \vertex (r) at (0, -2.2) {{\Large{\text{$(1\gamma^*,{\rm ISR})$}}}};
    \diagram* {
      (d) -- [fermion]  (u) -- [fermion] (a),
      (e) -- [charged scalar] (g) -- [charged scalar] (h),
      (g) -- [photon] (u),
      (p) -- [photon]  (u),
    };
\end{feynman}

\begin{scope}
\node[draw=black, circle, fill=white, minimum width=0.75cm, minimum height=0.75cm, xshift=-1cm] {};
\node[draw=black, circle, fill=1gISRlight, minimum width=0.75cm, minimum height=0.75cm, xshift=1cm] {};
\end{scope}
\end{tikzpicture}
\end{minipage}%
\hspace{0.01\textwidth}
\begin{minipage}{0.2\textwidth}
\begin{tikzpicture}[scale=0.5, transform shape]
\begin{feynman}[baseline=(a.b),small]
    \vertex (a) at (-2, -2) ;
    \vertex (b) at (-1, -1) ;
    \vertex (u) at (-1, 0) ;
    \vertex (p) at (-2, 0) ;
    \vertex (c) at (-1, 1) ;
    \vertex (d) at (-2, 2) ;
    \vertex [dot] (e) at (2, -2) ;
    \vertex (f) at (1, -1) ;
    \vertex (g) at (1, 0) ;
    \vertex [dot] (h) at (2, 2) ;
    \vertex (y) at (2,0) ;
    \vertex (r) at (0, -2.2) {{\Large{\text{$(1\gamma^*,{\rm FSR})$}}}};
    
    \diagram* {
      (d) -- [fermion]  (u) -- [fermion] (a),
      (e) -- [charged scalar] (g) -- [charged scalar] (h),
      (g) -- [photon] (u),
      (y) -- [photon]  (g),
    };
\end{feynman}

\begin{scope}
\node[draw=black, circle, fill=white, minimum width=0.75cm, minimum height=0.75cm, xshift=-1cm] {};
\node[draw=black, circle, fill=1gISRlight, minimum width=0.75cm, minimum height=0.75cm, xshift=1cm] {};
\end{scope}
\end{tikzpicture}
\end{minipage}%

\begin{center}
\vspace{-3mm}
\begin{tikzpicture}
\node (A) at (0,0) {};
\node (B) at (6,0) {}; 

\draw[decorate, decoration={brace, mirror, amplitude=10pt}]
(A) -- (B)
node[midway, below=12pt] {{\small{\text{$(1\gamma^*)$}}}};
\end{tikzpicture}
\end{center}

\begin{minipage}{0.2\textwidth}
\begin{tikzpicture}[scale=0.5, transform shape]
\begin{feynman}[baseline=(a.b),small]
    \vertex (a) at (-2, -2) ;
    \vertex (b) at (-1, -1) ;
    \vertex (u) at (-1, 0) ;
    \vertex (p) at (-2, 0) ;
    \vertex (c) at (-1, 1) ;
    \vertex (d) at (-2, 2) ;
    \vertex [dot] (e) at (2, -2) ;
    \vertex (f) at (1, -1) ;
    \vertex (g) at (1, 1) ;
    \vertex [dot] (h) at (2, 2) ;
    \vertex (y) at (0, -2.2) {{\Large{\text{$(2\gamma^*,{\rm ISR})$}}}};

    \diagram* {
      (d) -- [fermion] (c) -- [fermion] (u) -- [fermion] (b) -- [fermion] (a),
      (e) -- [charged scalar] (f) -- [charged scalar] (g) -- [charged scalar] (h),
      (g) -- [photon] (c),
      (b) -- [photon] (f),
      (p) -- [photon] (u),
    };
\end{feynman}

\begin{scope}
\node[draw=black, ellipse, minimum width=0.75cm, minimum height=3cm, xshift=-1cm, fill=white] {};
\node[draw=black, ellipse, minimum width=0.75cm, minimum height=3cm, xshift=1cm, fill=2gISRlight] {};
\end{scope}
\end{tikzpicture}
\end{minipage}%
\hspace{0.01\textwidth}
\begin{minipage}{0.2\textwidth}
\begin{tikzpicture}[scale=0.5, transform shape]
\begin{feynman}[baseline=(a.b),small]
    \vertex (a) at (-2, -2) ;
    \vertex (b) at (-1, -1) ;
    \vertex (u) at (1, 0) ;
    \vertex (p) at (2, 0) ;
    \vertex (c) at (-1, 1) ;
    \vertex (d) at (-2, 2) ;
    \vertex [dot] (e) at (2, -2) ;
    \vertex (f) at (1, -1) ;
    \vertex (g) at (1, 1) ;
    \vertex [dot] (h) at (2, 2) ;
    \vertex (y) at (0, -2.2) ;
    \vertex (y) at (0, -2.2) {\Large{\text{$(2\gamma^*,{\rm FSR})$}}};
    
    \diagram* {
      (d) -- [fermion] (c)-- [fermion] (b) -- [fermion] (a),
      (e) -- [charged scalar] (f) -- [charged scalar] (g) -- [charged scalar] (h),
      (g) -- [photon] (c),
      (b) -- [photon] (f),
      (p) -- [photon] (u),
    };
\end{feynman}

\begin{scope}
\node[draw=black, ellipse, minimum width=0.75cm, minimum height=3cm, xshift=1cm, fill=2gFSRlight] {};
\node[draw=black, ellipse, minimum width=0.75cm, minimum height=3cm, xshift=-1cm, fill=white] {};
\end{scope}
\end{tikzpicture}
\end{minipage}%
\begin{center}
\vspace{-3mm}
\begin{tikzpicture}
\node (A) at (0,0) {};
\node (B) at (6,0) {}; 

\draw[decorate, decoration={brace, mirror, amplitude=12pt}]
(A) -- (B)
node[midway, below=12pt] {{\small{\text{$(2\gamma^*)$}}}};
\end{tikzpicture}
\end{center}
\vspace{-4mm}
\caption{Topologies involved in the calculation of 
the structure-dependent corrections to $e^+e^-\to\pi^+\pi^-\gamma$. The notation $1\gamma^*$ and $2\gamma^*$ refers to the number of virtual photons connecting initial-state and final-state charged legs, while ISR and FSR are related to the signal photon.}
\label{fig:topologies_FsQED}
\end{figure}
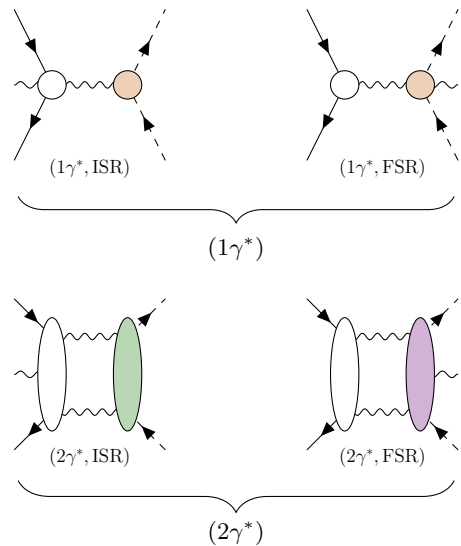
\section{One-loop calculation of structure-dependent corrections}
\label{sec:calculation}

In this section, we sketch the main features of the one-loop structure-dependent corrections to the $e^+ e^- \to \pi^+ \pi^-\gamma$ process. In the first subsection, we describe the calculation of virtual amplitudes introducing the pion form factor through its dispersive representation in loop integrals, both for final-state and initial-final state interference (IFI) contributions. In the second subsection, we discuss neutral pseudoscalar-pole contributions beyond FsQED in the second one. 

The full analytic amplitudes are derived with the \textsc{FeynRules}\cite{Alloul_2014}$\to$\textsc{FeynArts}~\cite{Hahn:1998yk,Hahn:2000kx,Hahn:2010zi}$\to$\textsc{FeynCalc}~\cite{Shtabovenko:2016sxi,Shtabovenko:2020gxv,Shtabovenko:2023idz} chain and are expressed as functions of one-loop scalar integrals. The loop functions are finally evaluated using the $\textsc{Collier}$~\cite{Denner:2014gla} library, with the exception of singular phase-space regions, where the amplitude is computed in quadruple precision using the \textsc{LoopTools}~\cite{Hahn:1998yk} package. The complete calculation is now available as part of the public release of the \textsc{BabaYaga@NLO} code through the project repository on \href{https://github.com/cm-cc/BabaYagaNLO}{GitHub \faGithub}.

\subsection{FsQED approach}
In the FsQED framework, the pion vector form factor can be described through a once-subtracted dispersion relation~\cite{Colangelo:2022lzg}, assuming elastic unitarity and the normalisation $F_\pi(0)=1$, as
\begin{equation}
F_\pi(q^2)=1+\frac{q^2}{\pi}\int_{4m_\pi^2}^{\infty}\frac{\dd s'}{s'}\frac{\Im F_\pi(s')}{s'-q^2-i\varepsilon'}\,,
\label{eq:ffdisp}
\end{equation}
together with the sum rule
\begin{equation}
\frac{1}{\pi}\int_{4m_\pi^2}^{\infty}\frac{\dd s'}{s'}\,\Im F_\pi(s')=1\,,
\label{eq:sumrule}
\end{equation}
which ensures $F_\pi(0)=1$ and implies $F_\pi(s)\to0$ for $s\to\infty$.
Since $F_\pi(0)=1$, the real corrections in the FsQED calculation coincide with those in F$\times$sQED. Differences arise only in the virtual contributions, where the dispersive representation generates structure-dependent terms in the one-loop amplitude. In particular, owing to the infrared behaviour of some loop diagrams, the combination $F_\pi(q^2)/q^2$ must be regularised by introducing a small fictitious photon mass $\lambda$, \textit{i.e.}
{\small \begin{equation}
\frac{F_\pi(q^2)}{q^2}\to
\frac{1}{q^2-\lambda^2+i\varepsilon'}
-\frac{1}{\pi}\int_{4m_\pi^2-\lambda^2}^{\infty}\frac{\dd s'}{s'}
\frac{\Im F_\pi(s'+\lambda^2)}
{q^2-s'-\lambda^2+i\varepsilon'}\,.
\label{eq:fregularised}
\end{equation}}

We adopt the same conventions as~\cite{CarloniCalame:2026vfc} and split the virtual FsQED amplitudes according to the number of virtual photons in the loop diagram, as
\begin{equation}
    \mathcal{A}_{\text{FsQED}}=2\Re\left\{\mathcal{A}_{\text{FsQED}}^{(1\gamma^*)}+\mathcal{A}_{\text{FsQED}}^{(2\gamma^*)}\right\}\,.
    \label{eq:fsqedrad}
\end{equation}

As pictured in Fig.~\ref{fig:topologies_FsQED}, four subsets can be further identified by distinguishing the current that emits the real photon in the virtual diagram.
The latter are associated to the hadronic processes $\gamma^* \gamma^* \to \pi^+\pi^-$  ($2\gamma^*,{\rm ISR}$), $i.e.$ the pion Compton tensor, and $\gamma^* \gamma^* \to \pi^+\pi^-\gamma$ ($2\gamma^*,{\rm FSR}$). Moreover, also corrections to the pion form factor ($1\gamma^*,{\rm ISR}$) and to  
the Compton tensor ($1\gamma^*,{\rm FSR}$) come into play. 
For the $\gamma^* \gamma^* \to \pi^+\pi^-$ topology, isolating the dominant contribution due to the pion pole is equivalent to FsQED~\cite{Colangelo:2022lzg}, while for $\gamma^* \to \pi^+\pi^-$ the full dispersive analysis is available~\cite{Monnard:2021pvm}. For radiative corrections to $\gamma^* \to \pi^+\pi^-\gamma$ and for the subprocess $\gamma^* \gamma^* \to \pi^+\pi^-\gamma$, dispersive analyses are still in progress~\cite{Aliberti:2024fpq,Hoferichter_Liverpool2025,Kaziukenas_Liverpool2025,Lymperiadou_Phipsi2026}. For all topologies, we assume in this work that FsQED is a good starting point to go beyond the standard F$\times$sQED approximation.

The amplitudes related to $(1\gamma^*)$ and $(2\gamma^*)$ contributions are detailed in Eqs.~\eqref{eq:fsr_fsqed} and~\eqref{eq:ifi_fsqed}. The kernels $\mathcal{A}^{(1\gamma^*,i)}_{j}(s')$ and $\mathcal{A}^{(2\gamma^*,i)}_{j}(s',s'')$ denote the interference between the tree-level and the renormalised virtual matrix elements in the on-shell scheme, as in~\cite{Budassi:2024whw,Budassi:2026lmr}. In this notation, the virtual photons are given squared masses $s'$ and $(s', s'')$, respectively, and the subscript $j$ indicates the current responsible for the tree-level photon emission.
In both expressions, the virtuality $\eta_i$ of the pion form factor is given by
\begin{equation}
    \eta_i = \begin{cases} M_{\pi\pi}^2 & \text{for $i =$ ISR,} \\ s & \text{for $i =$ FSR,} \end{cases}
\end{equation} where $M_{\pi\pi}$ is the di-pion invariant mass and $s$ is the squared c.m. energy.

As explicitly shown in Eq.~\eqref{eq:fsr_fsqed}, the infrared dependence of the $(1\gamma^*)$ contribution is entirely encoded in the first term $\mathcal{A}^{(1\gamma^*,i)}_{j}(\lambda^2)$. By contrast, the single- and double-dispersive integrals are finite, being free of both infrared and ultraviolet singularities (the latter after renormalization). Consequently, their numerical evaluation does not require any special treatment. Note that, in the double-dispersive integral, the coefficient of the product of the two imaginary parts comes from a partial fraction decomposition, using the same strategy as in~\cite{Budassi:2024whw}.

For the $(2\gamma^*)$ diagrams, whose complete expression is given in Eq.~\eqref{eq:ifi_fsqed}, the situation is significantly more subtle and demands a careful treatment. Unlike the $(1\gamma^*)$ case, the absence of a third pion form factor means that the infrared structure of these diagrams is not completely determined by the term $\mathcal{A}_{j}^{(2\gamma^*,i)}(\lambda^2,\lambda^2)$.
\begin{widetext}
\begin{equation}
\begin{aligned}
\hspace{-2.5cm}\mathcal{A}_{\text{FsQED}}^{(1\gamma^*)}
=&\sum_{i,j=\rm ISR,FSR}\biggl\{\mathcal{A}_{\text{FsQED},j}^{(1\gamma^*,i)}
\biggl\}F_\pi(\eta_i)F^*_\pi(\eta_j)\\
=&\sum_{i,j=\rm ISR,FSR}\biggl\{
\mathcal{A}^{(1\gamma^*,i)}_{j}(\lambda^2)
-\frac{2}{\pi}\int_{4m_\pi^2}^\infty\frac{\dd s'}{s'}\,\Im F_\pi(s') \mathcal{A}^{(1\gamma^*,i)}_{j}(s') \\
&+\frac{1}{\pi^2}\int_{4m_\pi^2}^\infty\frac{\dd s'}{s'}
\int_{4m_\pi^2}^\infty\frac{\dd s''}{s''}
\Im F_\pi(s')\Im F_\pi(s'') \frac{s''\mathcal{A}^{(1\gamma^*,i)}_{j}(s'')
-s'\mathcal{A}^{(1\gamma^*,i)}_{j}(s')}
{s''-s'-i\varepsilon''+i\varepsilon'}
\biggr\}
F_\pi(\eta_i)F^*_\pi(\eta_j)\,.
\end{aligned}
\label{eq:fsr_fsqed}
\end{equation}
\begin{equation}
\begin{alignedat}{2}
\hspace{0.4cm}\mathcal{A}_{\text{FsQED}}^{(2\gamma^*)}
&=\sum_{i,j=\rm ISR,FSR}\biggl\{\mathcal{A}_{\text{FsQED},j}^{(2\gamma^*,i)}\biggl\}F^*_\pi(\eta_j)\\
&=\sum_{i,j=\rm ISR,FSR}\biggl\{
\mathcal{A}^{(2\gamma^*,i)}_{j}(\lambda^2,\lambda^2)
&&-\frac{1}{\pi}\int_{4m_\pi^2}^\infty\frac{\dd s'}{s'}\,\Im F_\pi(s') \left[\mathcal{A}^{(2\gamma^*,i)}_{j}(\lambda^2,s')+\mathcal{A}^{(2\gamma^*,i)}_{j}(s',\lambda^2)\right] \\
& &&+\frac{1}{\pi^2}\int_{4m_\pi^2}^\infty\frac{\dd s'}{s'}
\int_{4m_\pi^2}^\infty\frac{\dd s''}{s''}
\Im F_\pi(s')\Im F_\pi(s'') \mathcal{A}^{(2\gamma^*,i)}_{j}(s',s'')
\biggr\}F^*_\pi(\eta_j)\,.
\end{alignedat}
\label{eq:ifi_fsqed}
\end{equation}
\end{widetext}

By explicitly separating the real part of each term in the sum in Eq.~\eqref{eq:ifi_fsqed}, we obtain
\begin{equation}
\begin{aligned}
\Re\biggl\{\mathcal{A}_{\text{FsQED},j}^{(2\gamma^*,i)}F^*_\pi(\eta_j)\biggl\}
= \biggl\{&
\Re F_\pi(\eta_j)\Re\mathcal{A}_{\text{FsQED},j}^{(2\gamma^*,i)} \\+ &\Im F_\pi(\eta_j)\Im\mathcal{A}_{\text{FsQED},j}^{(2\gamma^*,i)}
\biggr\}\,.
\label{eq:rereimim}
\end{aligned}
\end{equation}
It follows that the IR divergence is shared between the real and imaginary parts of the amplitude $\mathcal{A}_{\text{FsQED},j}^{(2\gamma^*,i)}$. In particular, the real part of the single-dispersive contribution of Eq.~\eqref{eq:ifi_fsqed} is IR sensitive. To address this issue and analytically isolate the IR-sensitive component of the dispersive integral, we exploit the fact that the IR singularity of the real part of the $\text{F}\times\text{sQED}$ amplitude is also preserved in FsQED, namely:
\begin{widetext}
\begin{equation}
    \left[\Re \mathcal{A}_{\text{F$\times$sQED},j}^{(2\gamma^*,i)}\right]_{\rm IR}\equiv\left[\Re\mathcal{A}^{(2\gamma^*,i)}_{j}(\lambda^2,\lambda^2)\,\Re F_\pi(\eta_i)\right]_{\rm IR} = \left[\Re \mathcal{A}_{\text{FsQED},j}^{(2\gamma^*,i)}\right]_{\rm IR}\,.
\end{equation}
\end{widetext}
Adding and subtracting in Eq.~\eqref{eq:ifi_fsqed} the structure $\left[\Re\mathcal{A}^{(2\gamma^*,i)}_{j}(\lambda^2,\lambda^2)\,\Re F_\pi(\eta_i)\right]$ allows to recast Eq.~\eqref{eq:ifi_fsqed} itself as in Eq.~\eqref{eq:realifi_fsqed}, where Eq.~\eqref{eq:ffdisp} has been used to rewrite $1 - \Re F_\pi(\eta_i)$. This procedure isolates the IR behaviour of the real part in the first term, while the single- and double-dispersive integrals remain IR finite.
\begin{widetext}
\begin{equation}
\begin{aligned}
\Re \mathcal{A}_{\text{FsQED},j}^{(2\gamma^*,i)}
=&
\Re \mathcal{A}^{(2\gamma^*,i)}_{j}(\lambda^2,\lambda^2)\Re F_\pi(\eta_i)\\
&-\frac{1}{\pi}\Re\int_{4m_\pi^2}^\infty\frac{\dd s'}{s'}\,\Im F_\pi(s') \left[\mathcal{A}^{(2\gamma^*,i)}_{j}(\lambda^2,s')+\mathcal{A}^{(2\gamma^*,i)}_{j}(s',\lambda^2)+\frac{\eta_i}{s'-\eta_i-i\varepsilon'}\Re \mathcal{A}^{(2\gamma^*,i)}_{j}(\lambda^2,\lambda^2)\right] \\
&+\frac{1}{\pi^2}\Re\int_{4m_\pi^2}^\infty\frac{\dd s'}{s'}
\int_{4m_\pi^2}^\infty\frac{\dd s''}{s''}
\Im F_\pi(s')\Im F_\pi(s'') \mathcal{A}^{(2\gamma^*,i)}_{j}(s',s'')\,.
\end{aligned}
\label{eq:realifi_fsqed}
\end{equation}
\end{widetext}
The single-dispersive contribution, however, still exhibits a pole at $s'=\eta_i$. The integration around this pole is treated using the $+i\varepsilon'$ prescription inherited from the dispersive integral. The second line of Eq.~\eqref{eq:realifi_fsqed} is therefore written as in Eq.~\eqref{eq:plemelji}, where the definition of $f(s')$ is given in Eq.~\eqref{eq:fplemelji} and $f(\eta_i^\pm)=\lim_{\delta\to 0^+} f(\eta_i\pm \delta)$. A thorough discussion of this relation is given in~\cite{Budassi:2024whw}. 
\begin{widetext}
    \begin{equation}
\begin{aligned}
  &\lim_{\varepsilon'\to 0_+}\Re\int {\dd s'} \frac{f(s')}{\eta_i-s'+i\varepsilon'} = \text{P.V.}\int {\dd s'}\left(\frac{\Re f(s')}{\eta_i-s'}\right) +\frac{\pi}{2} \Im f(\eta_i^+) +\frac{\pi}{2} \Im f(\eta_i^-)\, .\hfill \,
  \end{aligned}
\label{eq:plemelji}
\end{equation}
\begin{equation}
\begin{aligned}
& f(s') = -\frac{1}{\pi}\frac{\Im F_\pi(s')}{s'}\left[\mathcal{A}^{(2\gamma^*,i)}_{j}(\lambda^2,s')+\mathcal{A}^{(2\gamma^*,i)}_{j}(s',\lambda^2)+\frac{\eta_i}{s'-\eta_i-i\varepsilon'}\Re \mathcal{A}^{(2\gamma^*,i)}_{j}(\lambda^2,\lambda^2)\right](\eta_i-s'+i\varepsilon')\,.
\end{aligned}
\label{eq:fplemelji}
\end{equation}
\end{widetext}
The extraction of the infrared divergence associated with the imaginary part of $\mathcal{A}_{\text{FsQED},j}^{(2\gamma^*,i)}$ is more involved. In this case, the IR singularity is not contained in the imaginary part of the amplitude itself, but rather arises from the dispersive integration through an \textit{end-point singularity}, as pointed out in~\cite{Colangelo:2022lzg,Budassi:2024whw}. Compared to the case of two-pion production, the relevant topologies are more involved and include genuine five-point functions. To make use of the results of Ref.~\cite{Budassi:2024whw} concerning the dispersive integration of the imaginary part of the scalar four-point function, all five-point functions are reduced to combinations of four- and lower-point scalar functions using the package \texttt{hexagon.m}~\cite{Diakonidis:2008ij}. In this framework, only the scalar four-point integrals $D_0$ exhibit an end-point singularity in the limit $s' \to \eta_i$, whereas $A_0$, $B_0$, and $C_0$ remain regular for finite masses.

To isolate the IR divergent piece, to be dealt with analytically, and a finite term, which is integrated numerically, we first identify in the amplitude
$\mathcal{A}^{(2\gamma^*,i)}_{j}(\lambda^2,s')+\mathcal{A}^{(2\gamma^*,i)}_{j}(s',\lambda^2)$
all contributions proportional to scalar $D_0$ functions which exhibit the singular behaviour. We then add and subtract from the pole-dispersive integral the corresponding contribution, in which the coefficients of the singular $D_0$ terms are evaluated at $s'=\eta_i$ and multiplied by $\Im F_\pi(\eta_i)$. For brevity, we denote this subtraction term as
\begin{equation}
\Im F_\pi(\eta_i)\,\sum_l d_l(\eta_i)\, D_0^l(s')\,,
\end{equation}
where $d_l(\eta_i)$ represents the corresponding coefficients of the singular $D_0^l(s')$ evaluated at the end point.
With the above formulation, we can finally write the imaginary part of the amplitude of Eq.~\eqref{eq:rereimim} as
\begin{widetext}
\begin{equation}
\begin{aligned}
\Im \mathcal{A}_{\text{FsQED},j}^{(2\gamma^*,i)}
=&
\Im \mathcal{A}^{(2\gamma^*,i)}_{j}(\lambda^2,\lambda^2)-\frac{1}{\pi}\Im F_\pi(\eta_i)\sum_l d^l(\eta_i)\Im\int_{4m_\pi^2}^\infty\frac{\dd s'}{s'}\,D_0^l(s')\\&-\frac{1}{\pi}\Im \int_{4m_\pi^2}^\infty\frac{\dd s'}{s'}\,\left\{\Im F_\pi(s') \left[\mathcal{A}^{(2\gamma^*,i)}_{j}(\lambda^2,s')+\mathcal{A}^{(2\gamma^*,i)}_{j}(s',\lambda^2)\right]-\Im F_\pi(\eta_i)\sum_l d^l(\eta_i)D_0^l(s')\right\} \\
&+\frac{1}{\pi^2}\Im\int_{4m_\pi^2}^\infty\frac{\dd s'}{s'}
\int_{4m_\pi^2}^\infty\frac{\dd s''}{s''}
\Im F_\pi(s')\Im F_\pi(s'') \mathcal{A}^{(2\gamma^*,i)}_{j}(s',s'')\,.
\end{aligned}
\label{eq:imifi_fsqed}
\end{equation}
\end{widetext}
With this prescription, the kernel of the single-dispersive integral appearing in the second line of Eq.~\eqref{eq:imifi_fsqed} remains finite in the limit $s'\to\eta_i$, while the contribution associated with the end-point singularity, whose integral reproduces the correct IR behaviour, is isolated in the term
\begin{equation}
-\frac{1}{\pi}\Im F_\pi(\eta_i)\sum_l d^l(\eta_i)\Im\int_{4m_\pi^2}^\infty\frac{\dd s'}{s'}\,D_0^l(s')\,.
\label{eq:IRdispersive}
\end{equation}

This contribution requires special care. In contrast to the case of two-pion production, additional imaginary parts arise from cuts that are absent in~\cite{Colangelo:2022lzg,Budassi:2024whw}, since one of the external legs is off-shell in the present case. 
\begin{figure}[htbp]
\centering
\scalebox{0.8}{
\begin{tikzpicture}[decoration={brace, mirror, amplitude=8pt}]

    \begin{scope}
        \begin{feynman}[small]
            \vertex (a1);
            \vertex[right=1cm of a1] (b1);
            \vertex[below=1cm of a1] (i1);
            \vertex[below=1cm of b1] (j1);
            \vertex[above left=0.75cm and 0.75cm of a1] (c1);
            \vertex[below left=0.75cm and 0.75cm of i1] (d1);
            \vertex[above right=0.75cm and 0.75cm of b1] (e1);
            \vertex[below right=0.75cm and 0.75cm of j1] (f1);
            \vertex[above left=0.4 and 0.4 of a1] (g1);
            \vertex[below left=0.5 and 0.5 of g1] (h1);
            
            \diagram* {
                (a1) -- [graviton] (b1),
                (d1) -- [fermion] (i1) -- [fermion] (a1) -- [fermion] (c1),
                (e1) -- [charged scalar] (b1) -- [charged scalar] (j1) -- [charged scalar] (f1),
                (i1) -- [photon] (j1),
                (g1) -- [photon] (h1)
            };
        \end{feynman}
        \draw[red,dashed,thick] (0.5cm,0.75cm) -- (0.5cm,-1.75cm);
        
        \draw[decorate] ([yshift=-8pt]d1) -- ([yshift=-8pt]f1) node[midway, below=10pt] {\small $(2\gamma^*,\rm ISR)$};
    \end{scope}

    \begin{scope}[xshift=3.5cm]
        \begin{feynman}[small]
            \vertex (a2);
            \vertex[right=1cm of a2] (b2);
            \vertex[below=1cm of a2] (i2);
            \vertex[below=1cm of b2] (j2);
            \vertex[above left=0.75cm and 0.75cm of a2] (c2);
            \vertex[below left=0.75cm and 0.75cm of i2] (d2);
            \vertex[above right=0.75cm and 0.75cm of b2] (e2);
            \vertex[below right=0.75cm and 0.75cm of j2] (f2);
            \vertex[above right=0.4 and 0.4 of b2] (g2);
            \vertex[below right=0.5 and 0.5 of g2] (h2);
            
            \diagram* {
                (a2) -- [graviton] (b2),
                (d2) -- [fermion] (i2) -- [fermion] (a2) -- [fermion] (c2),
                (e2) -- [charged scalar] (b2) -- [charged scalar] (j2) -- [charged scalar] (f2),
                (i2) -- [photon] (j2),
                (g2) -- [photon] (h2)
            };
        \end{feynman}
        \draw[red,dashed,thick] (0.5cm,0.75cm) -- (0.5cm,-1.75cm);
    \end{scope}

    \begin{scope}[xshift=7cm]
        \begin{feynman}[small]
            \vertex (a3);
            \vertex[right=1cm of a3] (b3);
            \vertex[below=1cm of a3] (i3);
            \vertex[below=1cm of b3] (j3);
            \vertex[above left=0.75cm and 0.75cm of a3] (c3);
            \vertex[below left=0.75cm and 0.75cm of i3] (d3);
            \vertex[above right=0.75cm and 0.75cm of b3] (e3);
            \vertex[below right=0.75cm and 0.75cm of j3] (f3);
            \vertex[above right=0.4 and 0.4 of b3] (g3);
            \vertex[below right=0.5 and 0.5 of g3] (h3);
            
            \diagram* {
                (a3) -- [graviton] (b3),
                (d3) -- [fermion] (i3) -- [fermion] (a3) -- [fermion] (c3),
                (e3) -- [charged scalar] (b3) -- [charged scalar] (j3) -- [charged scalar] (f3),
                (i3) -- [photon] (j3),
                (g3) -- [photon] (h3)
            };
        \end{feynman}
        \draw[red, dashed, thick] (0.4, 0.5) to[out=-90, in=180] (1.5, -0.6);
    \end{scope}

    \draw[decorate] ([yshift=-8pt]d2) -- ([yshift=-8pt]f3) node[midway, below=10pt] {\small $(2\gamma^*,\rm FSR)$};

\end{tikzpicture}
}
\caption{Schematic representation of the cuts (red dashed lines) contributing to the imaginary part of the 4-point scalar function in $\mathcal{A}^{(2\gamma^*,\rm ISR)}_{j}(s',\lambda^2)$ and $\mathcal{A}^{(2\gamma^*,\rm FSR)}_{j}(s',\lambda^2)$. The double wavy line represents the massive $s'$ photon, whereas the single wavy line denotes the photon with vanishing mass $\lambda^2$.}
\label{fig:cutcosky}
\end{figure}
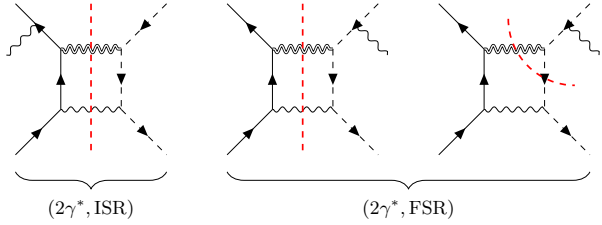

The cuts contributing to the imaginary part of the 4-point scalar function are shown in Fig.~\ref{fig:cutcosky}. For the FSR amplitudes $\mathcal{A}^{(2\gamma^*,\rm FSR)}_{j}(s',\lambda^2)$, the cut shown in the left panel generates an end-point singularity as $s' \to s$, leading to an IR-sensitive contribution after dispersive integration, while the alternative cut remains finite. For the ISR contributions $\mathcal{A}^{(2\gamma^*,\rm ISR)}_{j}(s',\lambda^2)$, only a single cut contributes, diverging in the $s'\to M_{\pi\pi}^2$ limit. Both singular structures reproduce the results previously derived in ~\cite{Colangelo:2022lzg,Budassi:2024whw}. It is worth noticing that under the interchange of $s'$ and $\lambda^2$, the imaginary part arising from each cut does not give rise to any end-point singularity, for both ISR and FSR virtual diagrams.
Accordingly, the dispersive integral associated with the singular imaginary part in Eq.~\eqref{eq:IRdispersive} is evaluated analytically, whereas the remaining imaginary contributions are computed numerically.
With the above theoretical framework, we obtain, for each phase-space point, the IR structure of the virtual amplitude that cancels the soft $\lambda$-dependent contribution. As a result, the dispersive integration is reconstructed by Monte Carlo sampling, generating for each phase-space point the virtualities $s'$ and $s''$.

\subsection{Neutral pseudoscalar-pole contributions}
\begin{figure*}[htbp] 
\begin{equation*}
    \begin{tikzpicture}[baseline={([yshift=-3pt]0,-1)}, scale=0.55, transform shape]
\begin{feynman}[small]
\vertex(a) at (0,0);
\vertex(b) at (0,-2);
\vertex(c) at (-2,0.25);
\vertex(d) at (-2,-2.25);
\vertex(e) at (2,0.25);
\vertex(f) at (2,-2.25);    
\vertex(g) at (0,-1);
\vertex(h) at (2,-1);
    \diagram* {
      (e)-- [charged scalar] (a) -- [charged scalar] (b) -- [charged scalar] (f),
(c) -- [photon] (a) ;
(d) --[photon] (b);
(g) --[photon] (h);  
    };
\end{feynman}
\begin{scope}
\node[draw=black, ellipse, minimum width=1.25cm, minimum height=3cm, xshift=0cm,yshift = -1cm, fill=2gFSRlight] {};
\end{scope}
\end{tikzpicture}
\qquad
\begin{tikzpicture}[baseline={([yshift=-3pt]0,-1)}]
    \node[inner sep=0pt] at (0,-0.5) {$=$};
\end{tikzpicture}
\qquad
    \begin{tikzpicture}[baseline={([yshift=-3pt]0,-1)}, scale=0.55, transform shape]
\begin{feynman}[small]
\vertex[blob](a) at (0,0) {};
\vertex[blob](b) at (0,-2) {};
\vertex(c) at (-2,0.25);
\vertex(d) at (-2,-2.25);
\vertex(e) at (2,0.25);
\vertex(f) at (2,-2.25);    
\vertex(g) at (0,-1);
\vertex(h) at (2,-1);
    \diagram* {
      (e)-- [charged scalar] (a) -- [charged scalar] (b) -- [charged scalar] (f),
(c) -- [photon] (a) ;
(d) --[photon] (b);
(g) --[photon] (h);  
    };
\end{feynman}
\end{tikzpicture}
\qquad\qquad
     \begin{tikzpicture}[baseline={([yshift=-3pt]0,-1)}, scale=0.55, transform shape]
\begin{feynman}[small]
\vertex[blob, fill=blue!50, pattern=](a) at (0,-1) {};
\vertex(c) at (-1.25,0.25);
\vertex(d) at (-1.25,-2.25);
\vertex[blob, fill=blue!50, pattern=](e) at (1.5,-1) {};
\vertex(f) at (2.5,-0);
\vertex(g) at (2.75,-2.25);
\vertex[above right=1.65 and 0 of f] (h);
\vertex[above right=0 and 1.65 of f] (i);
    \diagram* {
(c) -- [photon] (a) ;
(d) --[photon] (a);
(f) -- [graviton] (e);
(g) --[photon] (e);
(a) --[scalar, edge label={\Large $P$},inner sep=5pt] (e);
(h) --[charged scalar] (f);
(i)--[charged scalar] (f);
    };
\end{feynman}
\end{tikzpicture}
\qquad
\begin{tikzpicture}[baseline={([yshift=-3pt]0,-1)}, scale=0.55, transform shape]
\begin{feynman}[small]
\vertex[blob, fill=blue!50, pattern=](a) at (0,0) {};
\vertex[blob, fill=blue!50, pattern=](b) at (0,-2) {};
\vertex(c) at (-2,0.25);
\vertex(d) at (-2,-2.25);
\vertex(e) at (2,0.25);
\vertex(f) at (2,-2.25);
\vertex[above right=1.3 and 1.0 of e] (h);
\vertex[below right=1.0 and 1.3 of e] (i);
    \diagram* {
(c) -- [photon] (a);
(d) --[photon] (b);
(a) --[scalar, edge label={\Large $P$}] (b);
(f) -- [photon] (b);
(a) --[graviton] (e);
(h) --[charged scalar] (e);
(i)--[charged scalar] (e);
    };
\end{feynman}
\end{tikzpicture}
\qquad
\begin{tikzpicture}[baseline={([yshift=-3pt]0,-1)}, scale=0.55, transform shape]
\begin{feynman}[small]
\vertex[blob, fill=blue!50, pattern=](a) at (0,0) {};
\vertex[blob, fill=blue!50, pattern=](b) at (0,-2) {};
\vertex(c) at (-2,0.25);
\vertex(d) at (-2,-2.25);
\vertex(e) at (2,0.25);
\vertex(f) at (2,-2.25);
\vertex[above right=1.3 and 1.0 of e] (h);
\vertex[below right=1.0 and 1.3 of e] (i);
    \diagram* {
(d) -- [photon] (a);
(c) --[photon] (b);
(a) --[scalar, edge label={\Large $P$}] (b);
(f) -- [photon] (b);
(a) --[graviton] (e);
(h) --[charged scalar] (e);
(i)--[charged scalar] (e);
    };
\end{feynman}
\end{tikzpicture}
\end{equation*}
\caption{Representation of $(2\gamma^*,\rm FSR)$ contribution as given by the sum of the FsQED correction, where the shaded blob is the pion form factor $F_\pi(q^2)$, and the additional contributions are given by the $P=\pi^0,\eta,\eta^\prime$ pole in the HLbL, where the blue blob represents $F_{P\gamma^*\gamma^*}(q_1^2,q_2^2)$. Here the double wavy line represents the $\rho$ meson, where the corresponding TFF is amputated. The contraction with the initial $e^+e^-$ current is understood, with fixed external momenta.}
\label{fig:2gFSRpoles}
\end{figure*}
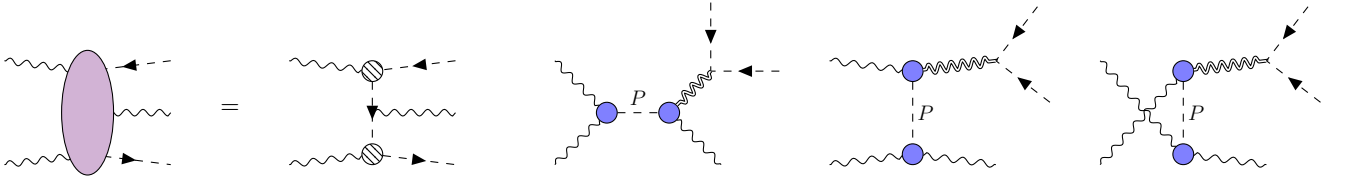

As emphasised in the literature~\cite{Aliberti:2024fpq,Hof_turin}, the $\gamma^*\gamma^*\to\pi^+\pi^-\gamma$ topology might receive possibly enhanced $P$-wave rescattering corrections that are not captured by the FsQED approach. As suggested in~\cite{Hof_turin,Hoferichter_private}, a first step to go beyond the FsQED calculation of the $(2\gamma^*,\rm FSR)$ contribution, is to consider the neutral pseudoscalar poles $P=\pi^0,\eta,\eta^\prime$ entering the Hadronic Light by Light (HLbL) scattering~\cite{Colangelo:2014dfa,Colangelo:2015ama}. We estimated such contributions by considering the $\gamma^*\gamma^*\to P \to\rho\,\gamma$ kernel, where the $\rho$ converts into a pion pair through the phenomenological coupling $\left|g_{\rho\pi\pi}\right|\simeq 6$~\cite{Klingl:1996by}. For the $\gamma^*\gamma^*P$ interactions, we estimate the transition form factor (TFF) $F_{P\gamma^*\gamma^*}(q_1^2,q_2^2)$ in a vector meson dominance (VMD) model~\cite{Knecht:2001qf,Hoferichter:2017ftn}, which represents an approximation of the full dispersive one~\cite{Hoferichter:2018kwz}, with the interaction vertex reading 
\begin{equation}
-i e^2\, \epsilon_{\mu\nu\alpha\beta}\, q_1^\alpha q_2^\beta F_{P \gamma^*\gamma^*}(q_1^2,q_2^2)\,.
\end{equation}
Due to the isospin conservation, the VMD amounts to inserting two massive propagators with the $\rho,\omega$ masses for the $\pi^0$ TFF, while for the $\eta^{(\prime)}$ two $\rho$ propagators are inserted, each with the appropriate $F_{P\gamma\gamma}=F_{P\gamma^*\gamma^*}(0,0)$ normalisation~\cite{Hoferichter:2018kwz,Holz:2024diw}. 
In order to get the correct amputated form factor for the $P\,\rho\,\gamma^*$ vertex, we substitute the TFF as~\cite{Hoferichter_private}
\begin{equation}
     F_{P \rho\gamma^*}(q_2^2)= \lim_{q_1^2\to M_\rho^2} F_{P \gamma^*\gamma^*}(q_1^2,q_2^2)\, \frac{g_{\rho\gamma}}{e M^2_\rho}\,(q_1^2-M_\rho^2)\,,
\end{equation} where $e$ 
is the electric charge and $|g_{\rho\gamma}|\simeq 5$~\cite{Hoferichter:2023mgy}. 

\section{Final-state radiation around the $\phi$ resonance}
\label{sec:radphi}

For a proper modelling of the $e^+ e^- \to \pi^+ \pi^-\gamma$ process at c.m. energies around the $\phi$-meson resonance, further mechanisms of real radiation beyond the F$\times$sQED approach need to be taken into account. Actually, these additional contributions to do not contribute to the definition of pion form factor and represent background processes.

As discussed in~\cite{Achasov:1997gb,Dubinsky:2004xv,
Czyz:2004nq, KLOE:2005jxf,Achasov:2005hm,Isidori:2006we,
Pancheri:2007xt,Gallegos:2009qu,Roca:2009zy,Shekhovtsova:2009yn,
PetitRosas:2026iuq}, the main relevant processes can be classified as follows: $a)$ direct radiative decays via a scalar resonance, $b)$ double resonance processes, and $c)$ bremsstrahlung processes.
The mechanisms $a)$ and $b)$ correspond to the processes
\begin{equation}
\begin{aligned}
    &a)\qquad e^+ e^- \to \gamma^* \to \phi \to S \, \gamma \to \pi^+ \pi^-\gamma \, ,\\
    &b)\qquad e^+ e^- \to V_1 \to (V_2 \, \pi) \to \pi^+ \pi^-\gamma \, ,
\end{aligned}
\label{eq:phiprocess}
\end{equation}
where $S = \{f_0, \sigma \}$ are the possible intermediate scalar mesons leading to the two-pion final state, while the vector resonances are given by $V_1= \{\phi,\omega'\}$ and $V_2=\rho^\pm$. Finally, while the contribution $c)$ inherently incorporates the baseline F$\times$sQED description of FSR, it extends this framework through the inclusion of additional bremsstrahlung processes involving vector and axial-vector mesons. For the energy region of interest, up to approximately 1 GeV, they can be modelled within a Chiral Perturbation Theory ($\chi$PT) framework containing the neutral $\rho^0$ meson, the charged $\pi^{\pm}$ and $a_1^{\pm}$ mesons, as well as the photon~\cite{Dubinsky:2004xv}.
Henceforth, to isolate these effects, we define contribution $c)$ as referring exclusively to these beyond-F$\times$sQED bremsstrahlung mechanisms.


As shown in~\cite{PetitRosas:2026iuq,Dubinsky:2004xv,Pancheri:2007xt}, the dynamics of all the above processes can be encoded into a FSR tensor, whose general form is dictated by Lorentz covariance and gauge invariance. In short-hand notation, the FSR tensor $M^{\mu\nu}$ describing the transition $\gamma^* \to \pi^+ \pi^-\gamma$ can be written as follows:
\begin{equation}
    M^{\mu\nu}_{\gamma^* \to \pi^+ \pi^-\gamma} \, = \, - i e^2 
    \, \sum_{i = 1}^{3} f_i \, \tau_i^{\mu\nu} \, ,
    \label{eq:fsrt}
\end{equation}
where $\tau_i^{\mu\nu}$ are three independent tensor structures as defined in~\cite{Dubinsky:2004xv} and $f_i$ are Lorentz-invariant functions. The decomposition 
of $M^{\mu\nu}_{\gamma^* \to \pi^+ \pi^-\gamma}$ as in 
Eq.~(\ref{eq:fsrt}) is model independent and the model dependence is only related to the explicit expression of the 
$f_i$ functions. Because of the distinct physical nature of the intermediate resonances involved in the FSR mechanisms $a)$, $b)$, and $c)$, each scalar function $f_i$ can be decomposed as:
\begin{equation}
    f_i \, = \, f_i^{V} + f_i^{S} + f_i^{\rm Br.} \, , \quad i = 1,2,3 \, ,
    \label{eq:fi_funct}
\end{equation}
where $f_i^{V}$, $f_i^{S}$, and $f_i^{\rm Br.}$ denote the vector, scalar, and beyond-F$\times$sQED bremsstrahlung contributions, respectively. In our calculation, the hadronic tensor of Eq.~\eqref{eq:fsrt} is contracted with the leptonic tensor of $e^+e^-\to\gamma^*$, squared and interfered with the F$\times$sQED amplitude.
\section{Numerical results}\label{sec:numres}

\begin{figure*}[htbp]
    \centering \includegraphics[width=0.95\textwidth]{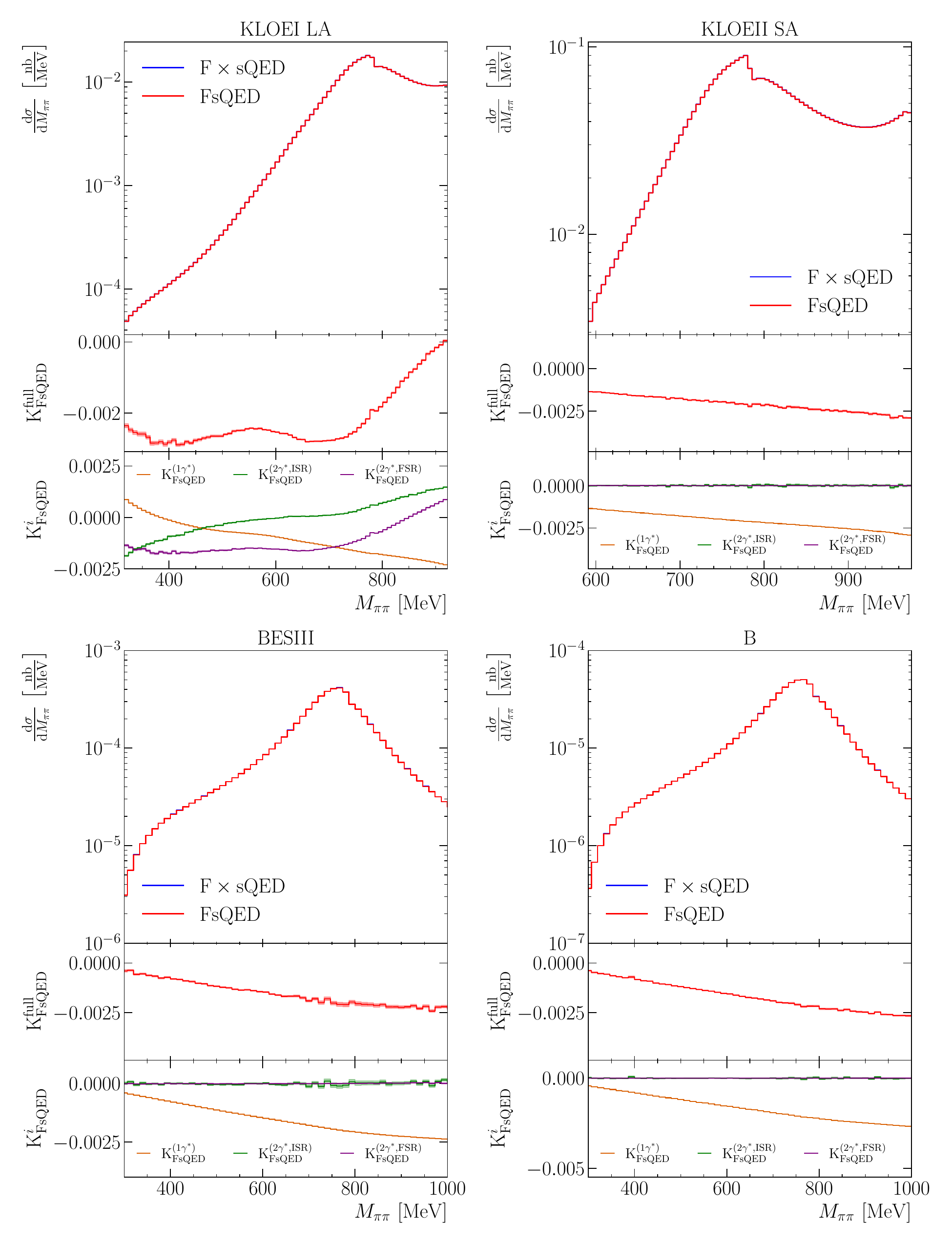}
    \caption{The differential cross section of the $e^+e^-\to\pi^+\pi^-\gamma$ process as a function of the invariant mass $M_{\pi\pi}$ in the four scenarios with the 
    F$\times$sQED and FsQED approaches. The ${\rm K}^i_{\rm FsQED}$ factors shown in the lower panels are defined in 
    Eq.~\eqref{eq:Kfactor}. The shaded area represent the $1\sigma$ statistical MC uncertainty.}
    \label{fig:invmass}
    \end{figure*}

\begin{figure*}[htbp]
    \centering
    \includegraphics[width=0.95\textwidth]{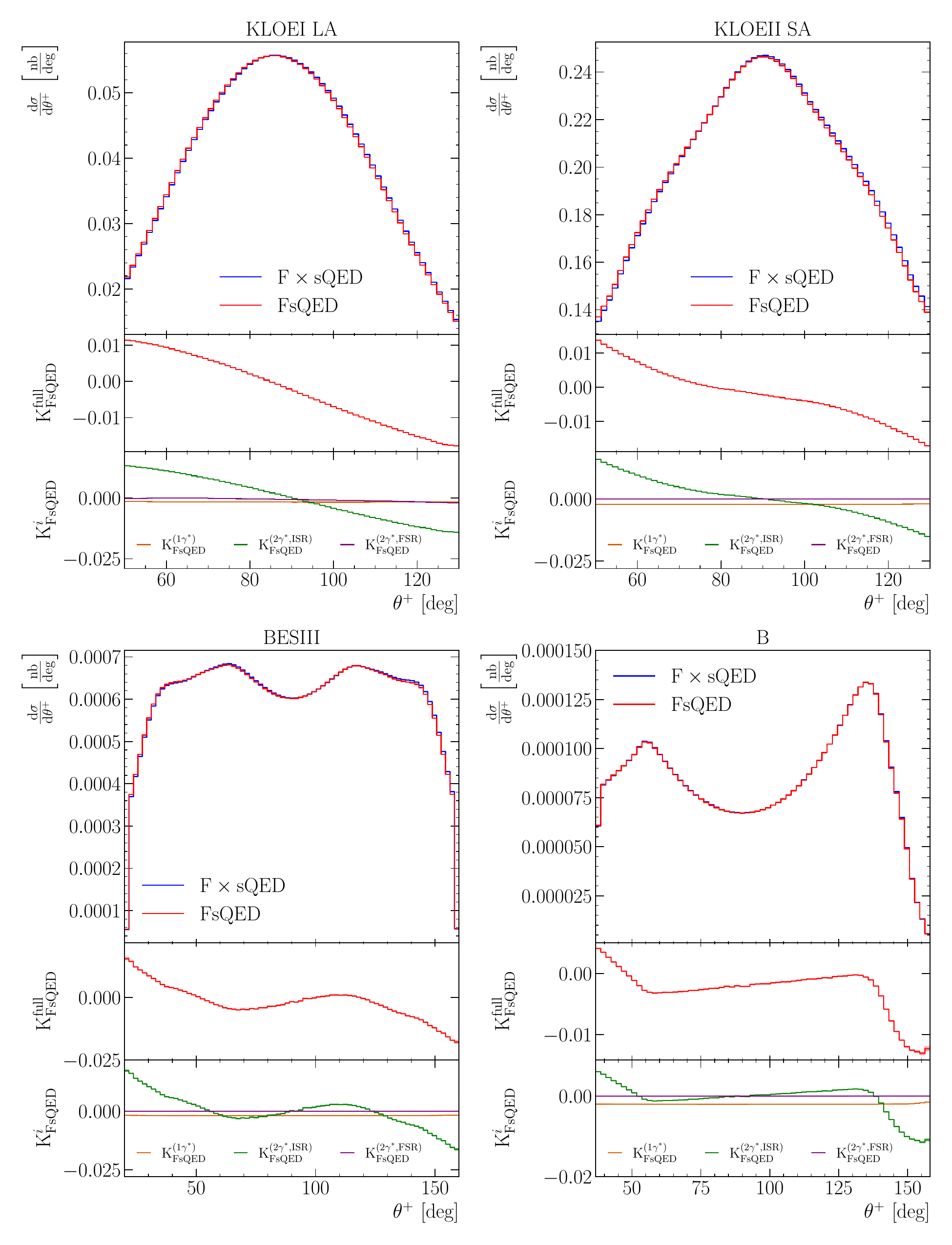}
    \caption{The same as in Fig.~\ref{fig:invmass} for the
    differential cross section as a function of the $\theta^+$ scattering angle.    
    }
    \label{fig:angdist}
\end{figure*}

We have implemented the NLO calculation described in Sec.~\ref{sec:calculation} in the Monte Carlo event generator \textsc{BabaYaga@NLO}~\cite{Budassi:2024whw,Budassi:2026lmr,Balossini:2006wc,Balossini:2008xr}. The FsQED fixed-order corrections have been matched with the Parton Shower (PS) algorithm reaching NLOPS accuracy, as done in~\cite{Budassi:2026lmr} for the factorised approach. Since the higher-order contributions are modified by sub-leading terms with respect to the $\rm F \times sQED$ case, only the structure-dependent NLO corrections are shown and discussed in this Section. Also the FSR mechanisms described in Sec.~\ref{sec:radphi} have been implemented in the code, dressing the LO amplitude with the PS algorithm.

For the numerical results, we use the following set of input parameters:
{\begin{equation}
    \begin{split}
    \alpha & = \, 1/137.03599908 \,, \\
     m_e & = \, 0.51099895~{\rm MeV}, \\
     m_{\pi}& =\, 139.57039~{\rm MeV}. 
\end{split}
\end{equation}}

For the $\rm F \times sQED$ and FsQED predictions, we employ the pion form factor given in~\cite{Aliberti:2024fpq}, while for the GVMD approach, we use the form factor parametrized as a sum of Breit-Wigner functions, as reported in~\cite{CarloniCalame:2026vfc}. Both form factors account for the six resonances $\rho$, $\omega$, $\phi$, $\rho^\prime$, $\rho^{\prime\prime}$ and $\rho^{\prime\prime\prime}$.
As previously noted, the scalar functions of Eq.~\eqref{eq:fi_funct} are inherently model dependent. In this work, both the specific models and their corresponding parameters are adopted directly from the literature. Specifically, the direct $\phi$ decay is parameterised according to~\cite{Achasov:2005hm}, the vector double resonant process follows~\cite{Pancheri:2007xt}, and the $\chi$PT
bremsstrahlung contributions are taken from~\cite{Dubinsky:2004xv}.

To model realistic event selection cuts used by radiative return experiments, we adopt the four scenarios described in~\cite{Aliberti:2024fpq} and also detailed in~\cite{Budassi:2026lmr}: KLOEI large-angle (LA) and KLOEII small-angle (SA) at c.m. energy of ${\sqrt{s}=1.02~{\rm GeV}}$, BESIII at ${\sqrt{s}=4~{\rm GeV}}$ and B at ${\sqrt{s}=10~{\rm GeV}}$.

\subsection{Structure-dependent corrections}
\label{sec:str_dep_num}
We study the impact of the structure-dependent corrections considering for all  scenarios the following differential cross sections:
\begin{equation}
    \frac{\rmd \sigma}{\rmd M_{\pi\pi}} \, ,
    \qquad \qquad \frac{\rmd \sigma}{\rmd \theta^+} \, ,
\end{equation}
where $M_{\pi\pi}$ is the di-pion invariant mass and $\theta^+$ is the polar scattering angle of the $\pi^+$ particle, defined with respect to the direction of the incoming electron. The former observable is the quantity used in the experiments to extract the pion form factor from data, while the latter is relevant for acceptance studies. In this Sec.~\ref{sec:str_dep_num}, the processes detailed in Sec.~\ref{sec:radphi} are omitted in order to isolate and evaluate precisely the pure structure-dependent corrections.

Our results for the two-pion invariant mass and the angular distribution are shown in Fig.~\ref{fig:invmass} and 
Fig.~\ref{fig:angdist}, respectively. The relative contributions shown in the lower panels of the plots are defined as follows:
\begin{equation}
    {\rm K}^i_{\rm FsQED} \, = \, 
    \frac{{\rmd \sigma}^i_{\rm FsQED} - {\rmd \sigma}^i_{\rm F\times sQED}}{{\rmd \sigma}_{\rm F\times sQED}} \, ,
    \label{eq:Kfactor}
\end{equation}
where ${\rmd \sigma}$ is a notation representing the differential cross section as a function of $M_{\pi\pi}$ or $\theta^+$. In Eq.~\eqref{eq:Kfactor}, 
the superscript $i$ is given by $i = {\rm full}, (1\gamma^*), (2\gamma^*,{\rm ISR}), (2\gamma^*,{\rm FSR})$, where ``full'' denotes the complete structure-dependent
correction. The separation of gauge-invariant contributions is performed at
the level of NLO amplitudes, which are interfered with the full
Born contribution.

As can be seen from Fig.~\ref{fig:invmass}, the complete set of FsQED corrections to the invariant mass around the $\rho$-peak is of the order of a few permille for all the experimental scenarios, from $\phi$- to $B$-factories. Across the three scenarios KLOEII SA, BESIII, and B, the $(1\gamma^*)$ subset represents the dominant contribution to ${\rm K}_{\rm FsQED}^{\rm full}$. For KLOEII SA, the NLO corrections are dominated by the $(1\gamma^*)$ contribution because of the enhancement driven by the squared collinear logarithm $L^2$, where $L=\log\!\left(s/m_e^2\right)$. 
On the other hand, in the BESIII and B cases, the $(2\gamma^*)$ contributions are strongly suppressed, as a consequence of the small value of $F_\pi(s)$. In the KLOEI LA scenario, all topologies contribute with a similar size to the full correction, due to the different photon angular cut with respect to KLOEII SA.

The structure-dependent corrections to ${\rmd \sigma} / 
{\rmd \theta^+}$ are at the percent level in all scenarios, as can be noticed from Fig.~\ref{fig:angdist}. In this case, the FsQED corrections are dominated by the $(2\gamma^*,\rm ISR)$ topology. This is due to the fact that the $(2\gamma^*,\rm ISR)$ diagrams contribute asymmetrically to $\rmd \sigma/\rmd \theta^+$ when interfered with the leading-order ISR, being enhanced by $|F_\pi(M_{\pi\pi}^2)|^2$ for $M_{\pi\pi}\simeq m_\rho$. 

It is worth noting that the numerical impact of the FsQED corrections as computed in this paper are in striking agreement with the effects discussed in~\cite{CarloniCalame:2026vfc} for the one-loop contributions to $e^+ e^- \to \pi^+ \pi^- \gamma$
in the GVMD model. Therefore, also for radiative return experiments, one observes an accordance between the predictions of the two formalisms for the computation of the structure-dependent corrections already noticed in~\cite{Budassi:2024whw} for the energy scan process $e^+ e^- \to \pi^+ \pi^-$.

\begin{figure}[t] 
\centering
\includegraphics[width=0.5\textwidth]{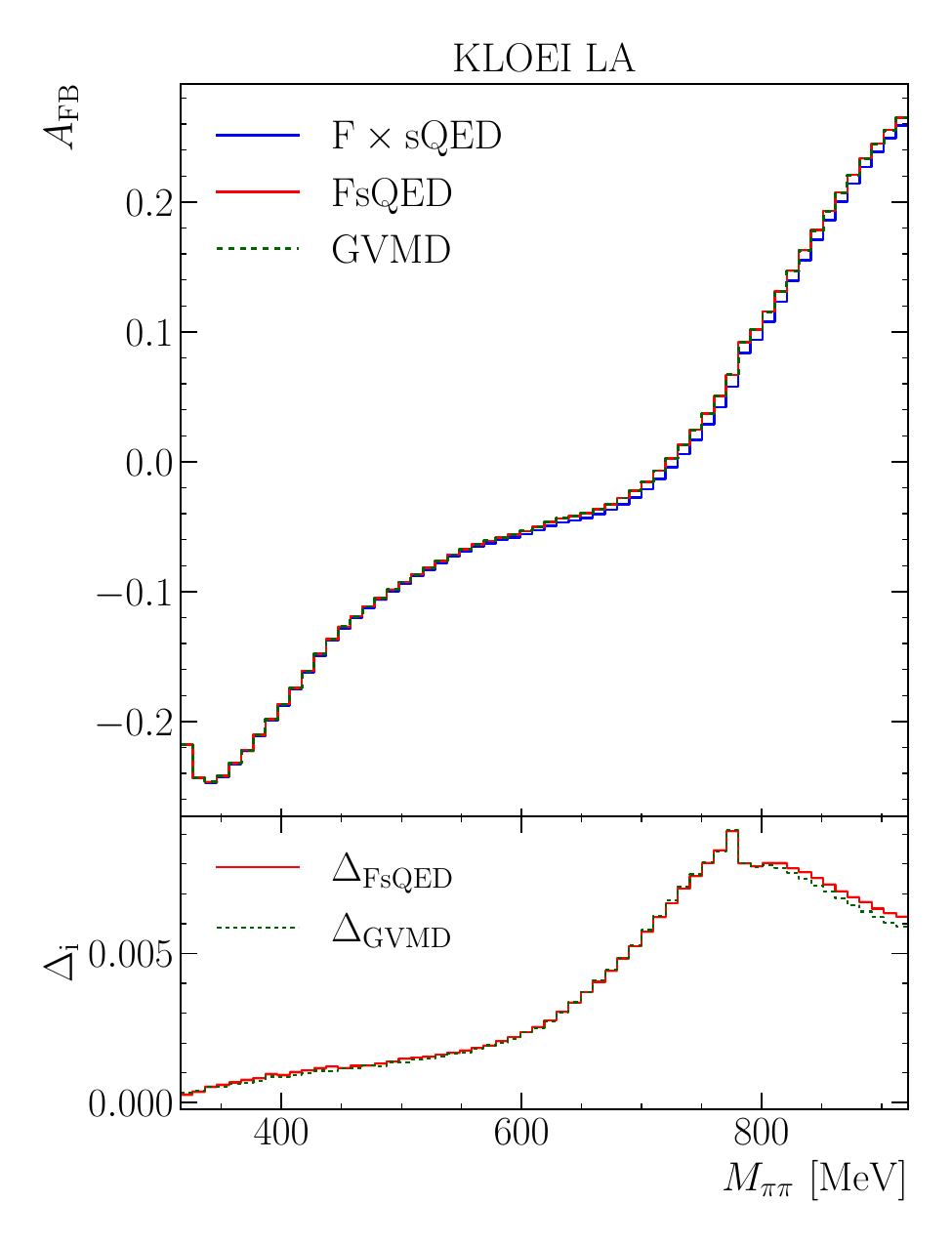} 
\caption{The forward-backward asymmetry of the $e^+ e^- \to 
\pi^+ \pi^- \gamma$ process in the KLOEI LA scenario. The absolute differences in the lower panel are defined in Eq.~\eqref{eq:deltaa}.}
\label{fig:asymm}
\end{figure}

 The additional contributions from 
neutral pseudoscalar poles in the $(2\gamma^*,\rm FSR)$ subset, depicted in Fig.~\ref{fig:2gFSRpoles}, remain below the permille level in all considered scenarios, both in the invariant mass spectrum $M_{\pi\pi}$ and in the angular distribution $\theta^+$. This can be understood as the $s$-channel diagram is helicity suppressed, whereas the $t$-channel and $u$-channel contributions, despite the behaviour of the $\rho$ propagator given by $(M_{\pi\pi}^2 - 
M_\rho^2 + i \Gamma_\rho M_\rho)^{-1}$, do not exhibit any infrared or collinear enhancement. A complete quantitative estimate, including beyond-FsQED effects other than pseudoscalar poles, deserves a more detailed analysis, which is left to a future investigation.

For completeness, we also study the size of FsQED corrections to the forward-backward asymmetry $A_{\rm FB}$ in the KLOEI LA setup, which is defined as:
\begin{equation}
    A_{\rm FB}
    \left(M_{\pi\pi}\right)=\frac{\rmd \sigma_{\rm F}-\rmd \sigma_{\rm B}}{\rmd \sigma_{\rm F}+\rmd \sigma_{\rm B}} \, ,
    \label{eq:asymmetry}
\end{equation}
where
\begin{equation}
\begin{split}
    \rmd \sigma_{\rm F}&=\int_{0}^{1}\frac{\rmd \sigma}{\rmd M_{\pi\pi}\,\,\rmd \cos{\theta^+}} \rmd \cos{\theta^+} \, ,\\
    \rmd \sigma_{\rm B}&=\int_{-1}^{0}\frac{\rmd \sigma}{\rmd M_{\pi\pi}\,\,\rmd \cos{\theta^+}} \rmd \cos{\theta^+} \, .
\end{split}
\end{equation} 
Our results for the forward-backward asymmetry are shown in 
Fig.~\ref{fig:asymm}. The absolute differences shown in the lower panel are defined as:
\begin{equation}
    \Delta_{\rm i} = A_{\rm FB}^{\rm i} - 
    A_{\rm FB}^{\rm F\times sQED} \qquad \rm i=\rm FsQED,GVMD
    \label{eq:deltaa}
\end{equation}
From Fig.~\ref{fig:asymm}, one can see that the full FsQED correction is about one percent close to the $\rho$-peak and it is at a few permille level below it. For comparison, we also display the prediction obtained within the GVMD approach of~\cite{CarloniCalame:2026vfc}. Remarkably, the two formulations exhibit an agreement at the $10^{-4}$ level, despite being based on completely different methodologies and pion form factor parameterisations.

Moreover, from Fig.~\ref{fig:asymm} it can also be noticed that the structure-dependent contributions to $A_{\rm FB}$ give rise to a resonant enhancement around the 
$\rho$, whose size is of the same order of magnitude as that due to the FsQED and GVMD corrections to the charge-asymmetry of the energy scan process 
$e^+ e^- \to \pi^+\pi^-$
\cite{Ignatov:2022iou,Colangelo:2022lzg,Budassi:2024whw,CarloniCalame:2026vfc}.

\begin{figure*}[t] 
\centering
\includegraphics[width=\textwidth]{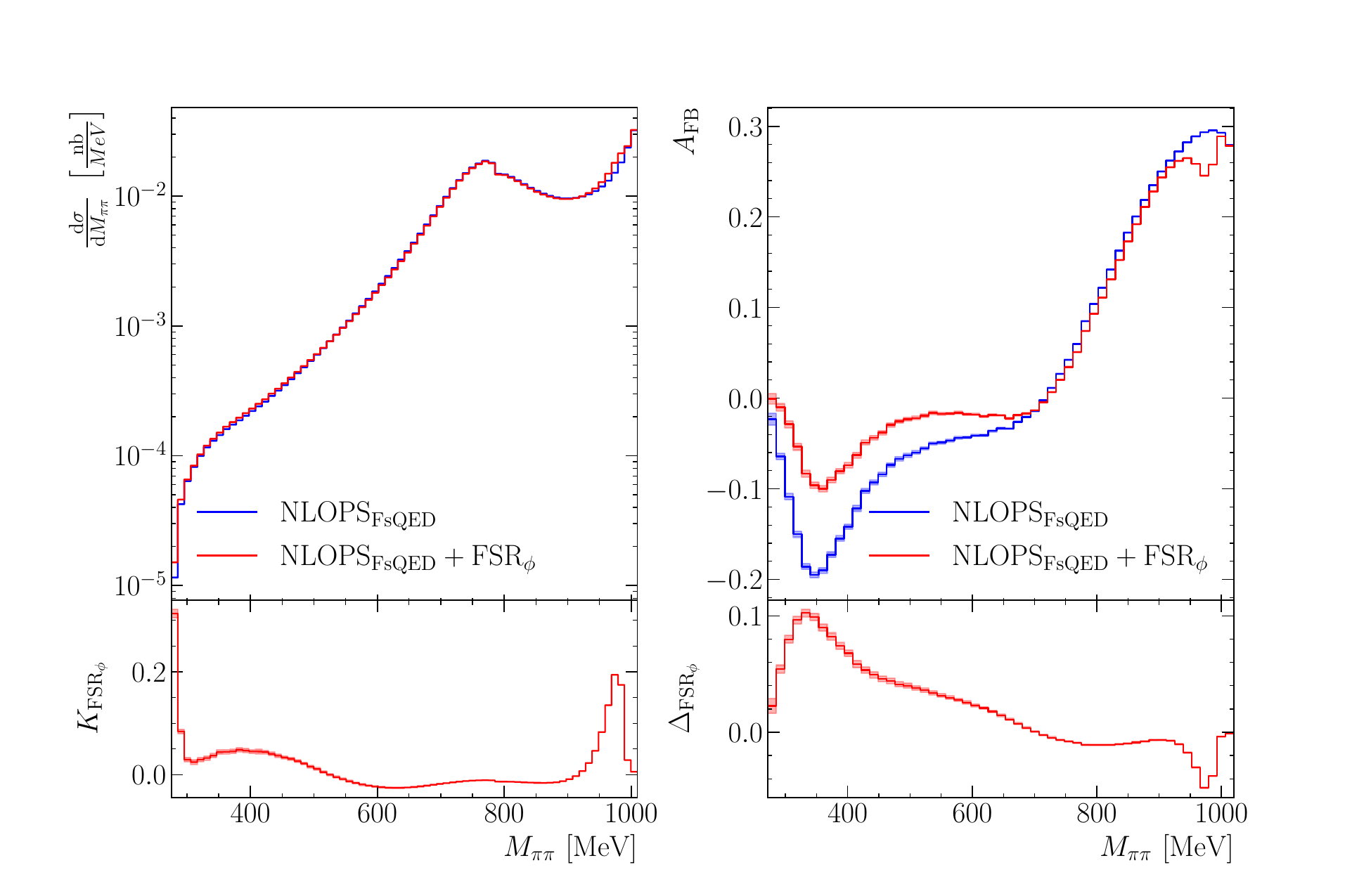}
\caption{Left panel: differential cross section for the $e^+e^-\to\pi^+\pi^-\gamma$ process as a function of the dipion invariant mass $M_{\pi\pi}$. Right panel: corresponding forward-backward asymmetry $A_{\rm FB}$. Blue curves represent the NLOPS prediction including structure-dependent FsQED corrections, while red curves show the impact of additionally incorporating the mechanisms discussed in Sec.~\ref{sec:radphi} at LOPS accuracy. The lower panels display the relative difference $K_{\rm FSR_\phi}$ normalized to the NLOPS$_{\rm FsQED}$ reference (left) and the absolute difference $\Delta_{\rm FSR_\phi}$ (right). The experimental selection cuts correspond to the scenario defined in Sec.~\ref{sec:pheno_fsrphi}.}
\label{fig:fsrphipheno}
\end{figure*}

\subsection{Final-state radiation around the $\phi$}
\label{sec:pheno_fsrphi}
In this Section, we show the impact of the different FSR processes 
discussed in Sec.~\ref{sec:radphi}. We first consider the radiative return setup at c.m. energy 
$\sqrt{s} = m_\phi$ of the KLOE experiment at DA$\Phi$NE, using a 
simplified event selection, namely:
\begin{equation}
\begin{split}
    &50^\circ \le \theta_{\pm} \le 130^\circ \, , \\
    &50^\circ \le \theta_{\gamma} \le 130^\circ \, , \quad E_\gamma \ge 10 \, \mathrm{MeV} \, .
\end{split}
\label{eq:setup_asymm}
\end{equation}
With this scenario, we can probe the invariant mass region 
up to approximately $1010 \, \mathrm{MeV}$, thus exploring the entire 
region around the $f_0$ resonance.
In Fig.~\ref{fig:fsrphipheno}, we investigate the role played by the mechanisms introduced in Sec.~\ref{sec:radphi} in a MC simulation of the two-pion invariant mass distribution and of the forward-backward asymmetry.

For both observables, we can note the effect of the $f_0$ resonance around 980 MeV induced by the direct radiative $\phi$ decay, that is of the order of several percent. Around the $\rho$ peak, the additional FSR mechanisms give small effects because of the dominance of the ISR contribution that is enhanced by $F_\pi(q^2)$ for $q^2\simeq m_\rho^2$. In the low invariant mass region, the $\phi$ radiative decay mediated by the $\sigma$ meson is significantly broadened due to the large width of the resonance, making it hardly visible. In the same region, double resonance mechanisms and $\chi$PT bremsstrahlung processes give a comparable contribution to the radiative $\phi$ decays. 

It is worth noting that, under the symmetric angular cuts of Eq.~\eqref{eq:setup_asymm}, the observables shown in Fig.~\ref{fig:fsrphipheno} provide complementary information on the additional FSR processes. Since the IFI exactly cancels out upon angular integration, the invariant mass distribution is sensitive only to even contributions under the exchange of the $\pi^+$ and $\pi^-$ momenta, thereby acting as a direct probe of the squared FSR amplitude. In contrast, the forward-backward asymmetry naturally isolates the odd terms, rendering the IFI the dominant contribution to this observable. As previously noted, the $f_0$ resonance significantly affects the shape of both observables. On the other hand, in the low invariant mass region, the $M_{\pi\pi}$ spectrum maintains its overall shape despite receiving large corrections, whereas the asymmetry undergoes a major change in this same kinematic range. 
Consequently, as already discussed in the literature~\cite{Pancheri:2007xt}, the near-threshold region of the asymmetry is highly sensitive to all FSR mechanisms, thus providing an important test for FSR dynamics. In particular, $\chi\text{PT}$ predictions can be tested independently using data away from the $\phi$-meson peak. 

\begin{figure}
    \centering
    \includegraphics[width=\linewidth]{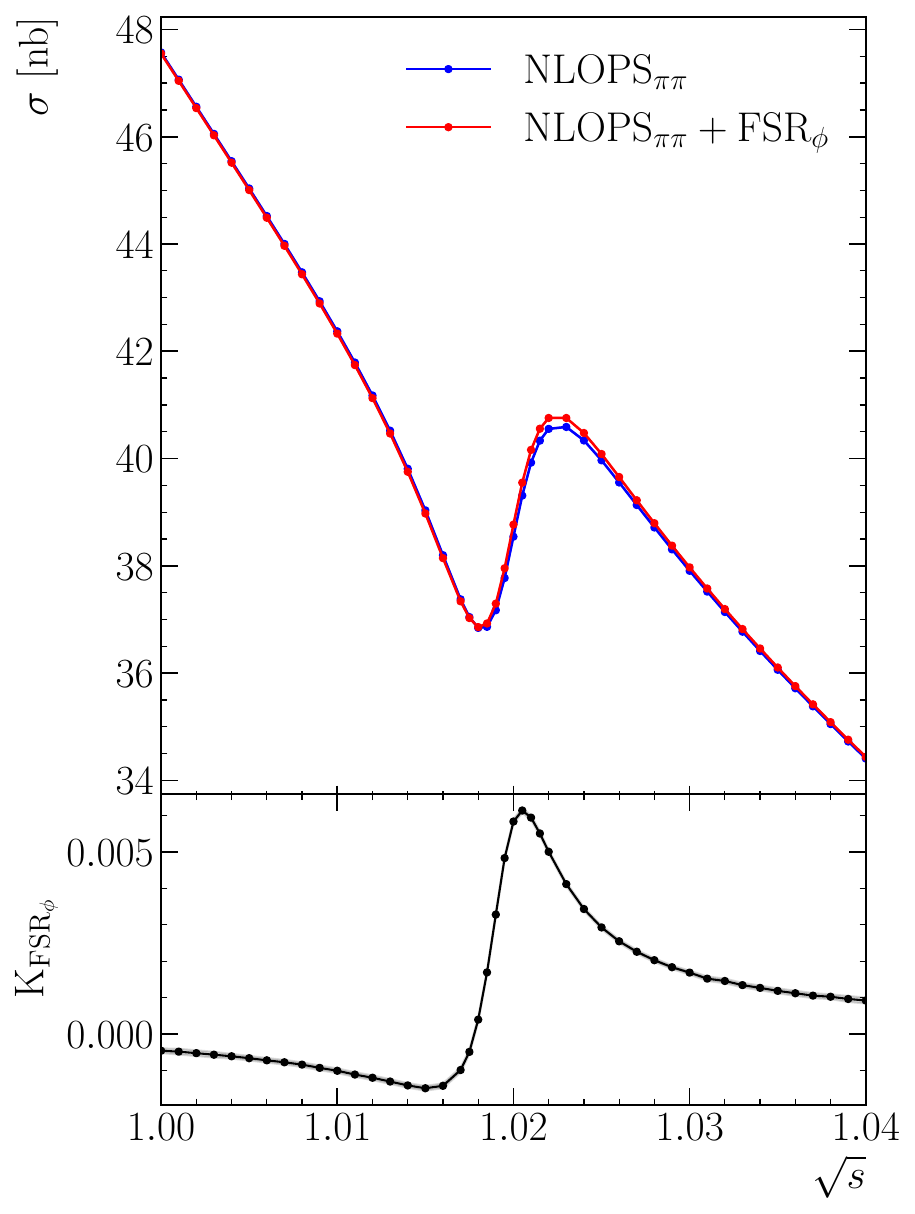}
    \caption{ Top panel: integrated cross section for the process $e^+e^-\to\pi^+\pi^-$ as a function of the c.m. energy $\sqrt{s}$ in the $\phi$-resonance region, computed at NLOPS accuracy excluding (blue curve) and including (red curve) additional FSR effects of Sec.~\ref{sec:radphi} at LOPS accuracy. Bottom panel: the corresponding relative difference normalised to the NLOPS result without additional FSR mechanisms.}
    \label{fig:scan_phi_cross_section}
\end{figure}

Furthermore, since the $e^+e^-\to \pi^+\pi^-\gamma$ channel constitutes a subset of radiative corrections to $\pi^+\pi^-$ production, we also evaluate the impact of these additional FSR mechanisms within an energy scan setup. For this purpose, we adopt the event selection criteria and the pion form factor parameterisation detailed in Ref.~\cite{Budassi:2024whw}, where these additional contributions were not investigated. Our numerical results indicate that all the corrections induced on the charge asymmetry remain strictly below the permille level across the entire energy range $\sqrt{s} < 1.2 \, \mathrm{GeV}$. Similarly, the relative effects on the integrated cross section are below the $10^{-3}$ level, with the exception of a narrow window around the $\phi$ resonance. In this region, as illustrated in Fig.~\ref{fig:scan_phi_cross_section}, the correction reaches up to $\sim 0.6\%$ and follows the expected resonant shape. Moreover, we verified that the bulk of this correction is driven by the direct radiative decay $a)$ of Eq.~\ref{eq:phiprocess}, which produces an enhancement in the differential cross section as a function of $M_{\pi\pi}$ around the $f_0$ mass region, similarly to what is observed in Fig.~\ref{fig:fsrphipheno} for radiative return. It is worth noticing that the $\chi$PT mechanisms are negligible in the entire region of interest for energy scan experiments.

\section{Conclusions}
\label{sec:conclusion}

We have computed the NLO structure-dependent corrections 
to the process $e^+ e^- \to \pi^+ \pi^- \gamma$, which is 
measured at flavour factories to extract the pion form factor 
in radiative return experiments. We have used a dispersive representation of the pion form factor
in the FsQED 
approach to include 
the composite structure of the pion in one-loop integrals associated with both FSR and IFI contributions.

We have compared the FsQED predictions with the results of the usual F$\times$sQED approximation for some observables of interest in radiative return measurements, from $\phi$- to $B$-factories. We have also discussed comparisons between the FsQED structure-dependent corrections and those obtained by us in~\cite{CarloniCalame:2026vfc} using the GVMD model.

We have shown that the invariant mass distribution is affected by FsQED corrections at the permille level in all experimental setup. 
On the other hand, the observables that are sensitive to the angular variables, such as 
the differential distribution of the scattering angle and the forward-backward asymmetry, receive corrections at the percent level. Moreover, we have emphasised that the FsQED corrections to radiative return are in remarkable agreement with the GVMD predictions of~\cite{CarloniCalame:2026vfc}, following the results of the energy scan scenario~\cite{Colangelo:2022lzg,Budassi:2024whw}.

According to recent input from the literature, we have also explored the role of pseudoscalar-pole contributions beyond FsQED and we have found that the numerical impact of such corrections is below the permille level in all the experimental scenarios. A more complete investigation of beyond-FsQED contributions is left to future work.

From our study, it is possible to conclude that the structure-dependent corrections to $e^+ e^- \to \pi^+ \pi^- \gamma$ are, in general, not negligible for radiative return measurements of the pion form factor at a sub-percent accuracy. 

In our work, we have also scrutinised the impact of the different mechanisms related to FSR that are known to be relevant for c.m. energies around the $\phi$ resonance. We have shown that these additional processes, namely the direct radiative decays mediated by the $f_0$ and $\sigma$ scalar resonances, the double resonance decays and the $\chi\text{PT}$ bremsstrahlung, provide significant contributions to the invariant-mass distribution and the forward-backward asymmetry at $\sqrt{s} = m_{\phi}$ and need to be taken into account for accurate MC simulations.

The FsQED computation presented in this work, as well as the 
improved treatment of FSR around the $\phi$-meson resonance, is available in a new version of the \textsc{BabaYaga@NLO} event generator, where also 
GVMD corrections to radiative return are implemented. Thanks to the results obtained 
in~\cite{Budassi:2024whw,Budassi:2026lmr,CarloniCalame:2026vfc}, in this release of 
\textsc{BabaYaga@NLO}, the structure-dependent corrections according to two independent approaches are matched to a multi-photon PS for both $e^+ e^- \to \pi^+ \pi^-$ and $e^+ e^- \to \pi^+ \pi^- \gamma$. This version of the code can be used to assess the uncertainty due to the modelling of the pion-photon interaction in both energy scan and radiative return experiments at flavour factories.

As a further prospect, we plan to compute the lepton pair corrections to the production of two pions in $e^+ e^-$ annihilation, in order to improve the accuracy in the calculation of higher-order contributions in \textsc{BabaYaga@NLO} for simulations and data analysis at flavour factories. We are also interested in extending our results on radiative corrections to hadron production processes in $e^+ e^-$ annihilation with predictions for the production of kaon pairs in energy scan and radiative return measurements.

\vskip 12pt\noindent
{\bf Data availability statement}
\vskip 8pt
The 
code used to produce all the numerical results shown in this paper is available from~\cite{babayaga@nlorep}.

\begin{acknowledgments}
We acknowledge Martin Hoferichter for valuable input on pseudoscalar-pole corrections beyond FsQED, as well as for reading a preliminary version of our manuscript and useful comments. We thank Gilberto Colangelo for helpful discussions. We are grateful to Stefan M\"uller for sharing information on the KLOE measurement of the forward-backward asymmetry.
 We would like also to thank Riccardo Aliberti, Achim Denig and Fedor Ignatov for general feedback and interest in our work.
AG acknowledges support by the Croatian Science Foundation~(HRZZ) project ``Beyond the Standard Model discovery and Standard Model precision at LHC Run~III'', IP-2022-10-2520.
\end{acknowledgments}

 \bibliography{FsQED}

\end{document}